\documentclass[trackchanges]{aastex701}

\usepackage[normalem]{ulem}
\usepackage{soul}
\usepackage{amsmath}
\usepackage{subcaption}
\usepackage{booktabs}
\usepackage{float} 
\usepackage{xcolor}
\begin{document}

\title{Constraining Orbital Eccentricity of a Supermassive Black Hole Binary Candidate  PKS 2131-021}

\author[orcid=0000-0002-8651-9510,sname='Paladi']{Avinash Kumar Paladi}
\affiliation{Joint Astronomy Programme, Department of Physics, Indian Institute of Science, Bengaluru 560012, India}
\email[show]{avinashkumarpaladi@gmail.com}

\author[orcid=0000-0003-4274-4369]{A. Gopakumar}
\affiliation{Department of Astronomy and Astrophysics, Tata Institute of Fundamental Research, Mumbai 400005, India}
\email[show]{gopu@tifr.res.in}

\author[orcid=0000-0001-5507-7660]{Sushmita Agarwal}
\affiliation{Department of Astronomy and Astrophysics, Tata Institute of Fundamental Research, Mumbai 400005, India}
\email[hide]{sush.agarwal16@gmail.com}

\author[orcid=0000-0003-2444-838X]{Fazal Kareem}
\affiliation{Max-Planck-Institut f{\"u}r Radioastronomie, Auf dem H{\"u}gel 69, 53121 Bonn, Germany}
\affiliation{Department of Physical Sciences, Indian Institute of Science Education and Research Kolkata 741246, India}
\email[hide]{fazalabdulkareem12@gmail.com}

\begin{abstract}
A detailed analysis of the decades-long radio light curve of blazar PKS 2131-021 showed epochs of sinusoidal variations in the radio flux density time-series as detailed in \cite{pks_rb}. 
The observed sinusoidal flux modulation arises naturally from relativistic Doppler boosting of the jet when the jet-emitting supermassive black hole (SMBH) orbits its companion.
For SMBHs in circular orbits, this scenario yields sinusoidal light curves, offering a simple kinematic explanation for the observed variability in  PKS 2131–021. 
We present an approach that incorporates the effects of orbital eccentricity into the Kinematic Orbital model for PKS 2131–021, using the Keplerian parametric solution to describe the SMBH binary orbit. Using the available radio light curve data, we demonstrate that the proposed SMBH binary likely possesses a residual orbital eccentricity,  which we constrain through detailed Bayesian parameter estimation studies to be $0.053\pm0.015$ with a Bayes factor of 3.15 over the circular model.
However, when the analysis accounts for the presence of red noise in the data using a Damped Random Walk (DRW) process, the circular model is preferred, giving an eccentricity upper limit of $e < 0.15$. 
Nevertheless, our efforts reveal that the Circular+DRW model is strongly favored. This model consistently recovers a coherent periodic signal across all datasets, with the orbital period remaining well-defined even when accounting for broader uncertainties. This analysis incorporated archival observations from the Haystack Observatory, the University of Michigan Radio Astronomy Observatory (UMRAO), and the Owens Valley Radio Observatory (OVRO), spanning the period from 1975 to 2021, compiled by \cite{pks_rb}. 

\end{abstract}


\section{Introduction} \label{sec:intro}

The International Pulsar Timing Array (IPTA) consortium that employs data from an array of radio milli-second pulsars (MSPs)  is expected to establish the era of nano-Hertz (nHz) gravitational wave (GW) astronomy in the coming years \citep{ipta3Pc}. This is mainly because of the recently reported strong evidence for a persistent nHz GW background by the IPTA constituents in America, Australia, China, Europe, India and South Africa \citep{aaa+2023b,rzs+2023,xcg+2023,epta+inpta2023a,MPTAgwb}. A cosmological population of inspiraling Supermassive Black Hole (SMBH) binaries with milli-parsec orbital separations are most likely sources for such a nHz GW background, though there exists other plausible explanations \citep{aaa+2023c, epta+inpta2024}. Identifying the presence of such SMBH binaries with the help of electromagnetic observations should provide additional confidence to the astrophysical interpretation of the reported presence of nHz GWs in the rapidly maturing PTA data sets \citep{nggwsmbhb,eptagwsmbhb}. More importantly, individual SMBH binaries with milli-parsec orbital separations are potential sources to pursue persistent multi-messenger nHz GW astronomy in the era of  Square Kilometre Array (SKA)-based PTAs \citep{jgp+2022,WangMohantySKAPTA,PadmanabhanLoebSKA}.

\par
Several potential SMBH binary candidates have milli-parsec orbital separations  \citep{obssmbhb}. However, there are only two strong candidates, namely blazars OJ~287 \citep{OJ287Voltonendey2021} and PKS 2131-021 \citep{pks_rb}. The former exhibits quasi-periodic optical outbursts roughly every 12 years, and superposed on these optical outbursts are certain flares occurring in predictable double peaks. In the SMBH binary central engine description for OJ~287, these flares arise when a smaller secondary SMBH plunges through the accretion disk of a more massive primary SMBH twice every orbit, causing thermal flares whose epochs could be predicted accurately \citep{Dey2018, Dey2019}. In contrast, the SMBH binary description for  PKS 2131-021 arises as its radio light curve shows epochs of remarkably periodic sinusoidal variations which could be naturally explained by Doppler boosting from a jet tied to the primary black hole in a tight binary orbit with a roughly two-year period \citep{pks_rb, PKS2025}. A critical requirement for both these candidates is the availability of decades-long observational data \citep{obssmbhb}. In a recent targeted search using the NANOGrav 15-year dataset, \cite{Agarwal2025CGW} investigated continuous gravitational wave (CGW) emission from these sources. They reported stringent $95\%$ upper limits on the gravitational wave strain of $h_0 < 1 \times 10^{-14}$ and constrained the chirp masses to $\mathcal{M} < 1.5 \times 10^{10}\,M_{\odot}$ for PKS 2131-021.

\par 
The radio lightcurve of PKS 2131-021 shows a periodic sinusoidal signal observed in early Haystack data (1975–1983), which re-emerged after a 19-year quiescent phase in the more recent OVRO data (2008–2021), maintaining consistent phase and period \citep{pks_rb}. The SMBH binary central engine description for  PKS 2131-021, detailed in \cite{pks_rb}, argued that these observed sinusoidal flux-density variations, with a period of $2.082 \pm 0.003$ years in the rest frame of the source, naturally arise from the orbital motion of the SMBH that powers the observed radio jet. Their model requires that the observed blazar jet originates from the more massive primary SMBH, and the orbital motion of the less massive secondary SMBH induces a periodic wobble in the observed relativistic jet. This produces modulations in the Doppler boosting of the radio jet, leading to the observed sinusoidal brightness fluctuations in radio frequencies from blazar PKS 2131-021. Moreover, extensive red-noise simulations robustly rule out stochastic variability inherent in blazars as the origin of the observed signal. Specifically, the hypothesis that the variations arise from random fluctuations in the red-noise tail of the power spectral density is rejected with a p-value of $1.58 \times 10^{-6}$.

\par 
A recent effort, detailed in \cite{PKS2025},  further substantiated the conclusions of \cite{pks_rb} by identifying a coherent, sinusoidal periodic variation in the recent OVRO 15 GHz light curve, extended through 2023, the dataset spans a remarkable 47.9-year baseline, providing an unprecedented temporal window to study long-term variability in PKS~2131–021. This extended dataset rejects the observed sinusoidal signal to be due to red noise with a p-value of $2.09 \times 10^{-7}$, strengthening its periodic/sinusoidal nature. Further, a similar periodic modulation is seen in multi-wavelength data from radio (up to 345 GHz from ALMA) and in optical observations from ZTF, with a systematic phase shift indicating that higher-frequency emissions lead lower-frequency ones \citep{PKS2025,pks2131ren}. The gamma ray light curve from Fermi LAT also show flares consistent with the observed periodicity \citep{PKS2025}. Interestingly, a similar sinusoidal signal is also reported in the Atacama Cosmology Telescope (ACT) 95, 147 and 225 GHz lightcurves of PKS 2131-021 \citep{pks2131act}. These compelling multi-band observational and statistical inferences indicate a persistent, underlying clock-like mechanism, naturally associated with an SMBH binary description.

\par
We observe that \cite{pks_rb} let the constituents SMBHs move in circular orbits. We provide a computationally efficient prescription to extend the existing SMBH binary central engine description for PKS 2131-021 by allowing the SMBHs to move in eccentric orbits. We let the SMBHs move in tiny eccentricity orbits as the observed periodic signal remains nearly sinusoidal, suggesting the underlying binary has only a small orbital eccentricity. We adapt inputs from a widely used \texttt{ELL1} timing model, employed to time pulsar binaries with nearly circular orbits \citep{ELL1}. Specifically, we utilize the Laplace-Lagrange parameters, $\epsilon_1 \equiv e\sin\omega$ and $\epsilon_2 \equiv e\cos\omega$, to construct an analytical ELL1 model for SMBH binaries with tiny eccentricities. 
We subsequently perform Bayesian parameter estimation and model selection using the observational data compiled by \cite{pks_rb}, and this allows us to directly compare the circular orbit model against our eccentric SMBH binary descriptions with and without the inclusion of stochastic red noise. 
Our results indicate that while the data favor a small but non-zero residual eccentricity in the absence of red noise, the preference shifts to a circular model when a Damped Random Walk (DRW) process is incorporated. In the latter case, we establish a $95\%$ upper limit on the eccentricity of $e < 0.15$, and  discuss the implications of our efforts even for the circular model.

\par
This paper is structured as follows. The next section provides a brief review of the circular SMBH binary model of \cite{pks_rb} and this is followed by a detailed derivation of our approach to bring in the effects of orbital eccentricity into the exiting Kinematic Orbital model for PKS 2131-021. A brief description of the employed radio light curve for PKS 2131-021 is provided in Section \ref{sec:obsevid}, followed by the application and discussion of our eccentric models, both with and without the inclusion of red noise.
We summarize our results and discuss avenues for possible improvements in section \ref{sec:conclusions}.

\section{ SMBH Binary  Descriptions for PKS 2131-021} 
\label{sec:SMBHBmodel}

We begin by summarizing the SMBH binary model of \cite{pks_rb} where SMBHs move in quasi-circular orbits. How we incorporate orbital eccentricity into their quasi-circular orbital description is detailed in Section.~\ref{subsec:ell1}.

\subsection{Circular SMBH Binary Model of \cite{pks_rb}} 
\label{subsec:cir}

To explain the sinusoidal nature of the radio light curve, observed around 15 GHz, \cite{pks_rb} invoked relativistic Doppler boosting due to the orbital motion of the jet emitting SMBH in  PKS 2131-021. In other words, the jet-launching SMBH's orbital motion causes periodic changes in the angle of the jet relative to us. Because relativistic Doppler boosting is highly sensitive to this angle, even small orbital movements could cause strong, periodic modulations in the observed jet emission, as seen in PKS 2131-021. This scenario should be relevant for PKS 2131-021 due to the reported observations of superluminal motion in this blazar, which requires a very small angle between its radio jet and our line of sight, estimated to be $ \sim 5.7^\circ$ and the estimated high Doppler factor of its jet material \citep{Lister2019,Homan2021}.

\par
\begin{figure}[ht!]
\centering
\includegraphics[width=\linewidth]{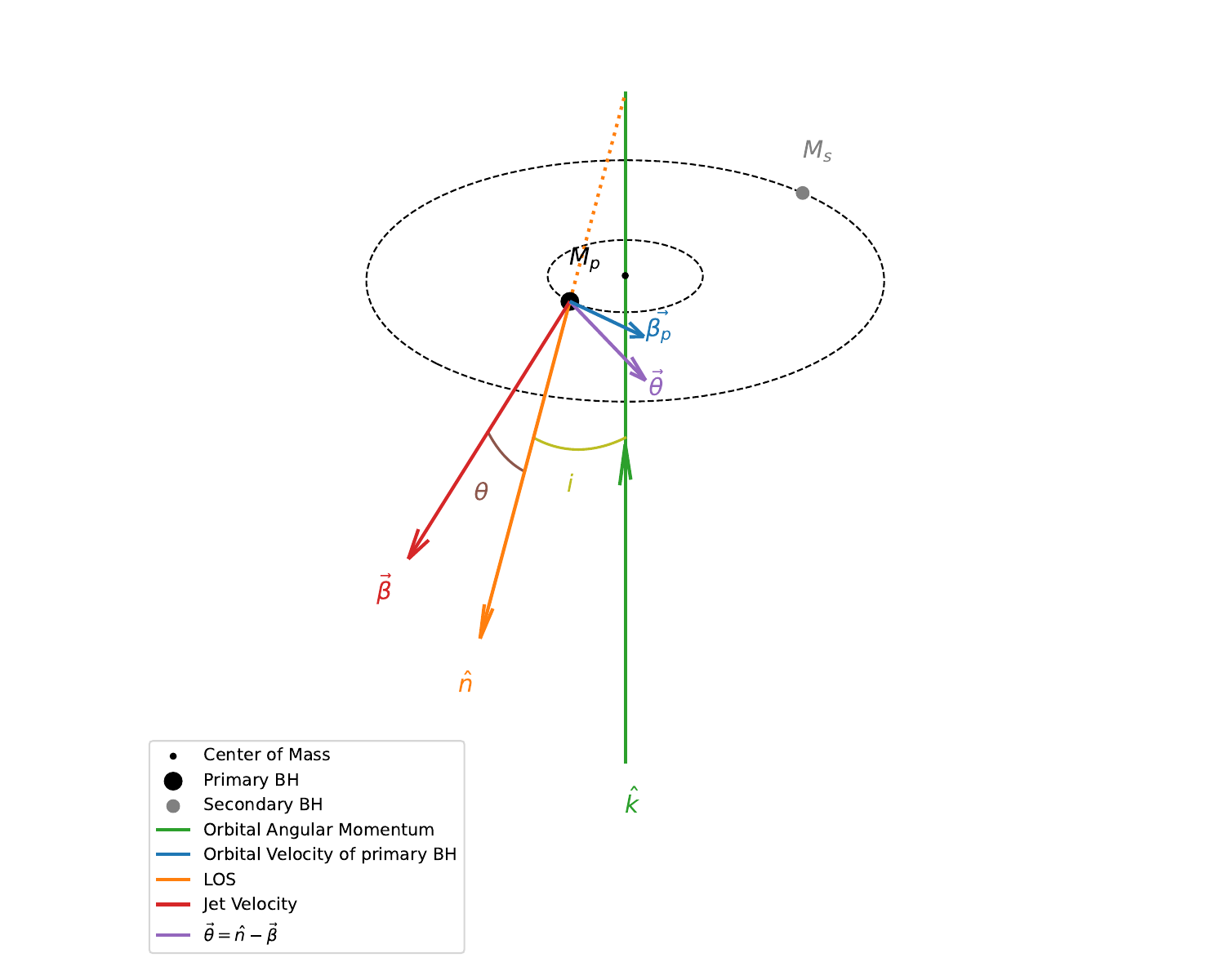}
\caption{Model explaining the sinusoidal flux density variations in Blazar PKS 2131-021. The primary SMBH with mass $M_p$ and the secondary SMBH with mass $M_s$ revolve around the common center of mass in a circular orbit with velocities $c\vec{\beta_p}$ and $c\vec{\beta_s}$. The orbital angular momentum vector $\hat{k}$ is shown perpendicular to the plane of the circular orbit, which is inclined at an angle $\iota$ to the line of sight ($\hat{n}$). The jet from the primary SMBH with velocity $c\vec{\beta}$ is oriented at a small angle $\theta$ from the line of sight vector $\hat{n}$. $\vec\theta = \hat{n} - \vec{\beta}$ is a vector in the plane of the sky or orthogonal to the line of sight ($\hat{n}$).}
\label{plt:cartoon}
\end{figure}

 
Influenced by \cite{pks_rb,PKS2025}, we display Fig.~\ref{plt:cartoon} to define various quantities that are required to describe the SMBH binary description for PKS 2131-021. This scenario involves two SMBHs of masses $M_p$ and $M_s$ that move around each other with an orbital period of $P_b$ in the binary rest frame. Further, it is convenient to let the jet originate from  ${\rm M_p}$, and we assume that the relativistic jet is launched along the SMBH spin axis with constant velocity $c{\vec\beta}$ relative to $M_p$. The fact that this is a BL Lac object ensures that the line of sight is inclined at a small angle $\theta$ to the jet axis. Further, we let $\iota$ be the angle between the observer's line of sight $( \hat{n})$ and SMBH binary orbital angular momentum $(\vec{k})$, and let $ \vec{\beta_p}$ denote the orbital velocity of $M_p$. It is important to introduce a vector $\vec\theta = \hat{n} - \vec{\beta}$, which is in the plane of the sky and therefore perpendicular to $\hat{n}$. The orbital velocity of the primary $\beta_p$ can be expressed in terms of $M_p$ and $M_s$ (expressed in $M_{\odot}$) as follows:

\begin{equation}
\beta_p= \left(\frac{2\pi T_{\odot}}{P_b} \right)^{\frac{1}{3}} \frac{M_s}{(M_p +M_s)^\frac{2}{3}}\,,
\label{eq:doppler}
\end{equation}
where $T_\odot = GM_\odot/c^3$. 
We note in passing that although the jet is assumed to originate from the primary SMBH, the kinematical model remains valid for the case where the jet is emitted from the secondary SMBH.

\par
In blazars like PKS 2131-021, the relativistic jet points nearly toward Earth, causing strong flux density boosting due to Doppler beaming, and it dramatically amplifies the observed brightness, shifts the spectrum, and compresses timescales, ensuring that blazars appear unusually bright, variable, and energetic \citep{UP95}. For an intrinsically isotropic emitter with rest-frame flux density $S'$, the observed flux density $S$ is Doppler-boosted according to the relation \citep{1979Natur.277..182S}:
\begin{equation}
S= {\cal D} ^{2-\alpha}S'\,,
\label{eq:doppler}
\end{equation}
where the spectral index $\alpha$  and the Doppler factor
${\cal D}$ are given by 
\begin{subequations}
\begin{align}
\alpha &=\frac{d\ln S}{d\ln f},,
\\
{\cal D} &=\frac1{\gamma(1-{\vec\beta}\cdot{\hat{n}})}\,,
\end{align}
\end{subequations}
where $f$ is the observational frequency and $\gamma=(1-\beta^2)^{-1/2}$ is indeed the Lorentz factor. It is important to note that the orbital motion of $M_p$ causes both $\gamma$ and $\vec\beta$ to vary with time.

\par 
To model the observed varying flux density, we employ Eq.~\ref{eq:doppler} and write the fractional change in observed flux as 
\begin{align}
\delta\ln S&=(2-\alpha) \delta \ln {\cal D}=(2-\alpha)\delta{\vec\beta}\cdot\frac{\partial\ln {\cal D}}{\partial{\vec\beta}}
\nonumber  \\
&=\frac{(2-\alpha)({\hat{n}}-{\vec\beta})\cdot{\vec\beta}_{p}}{(1-{\hat{n}}\cdot{\vec\beta})}+\mathcal{O}(\beta_{p}^2)\,,
\label{eq:fluxchange}
\end{align}
where we let $\hat{n}$ be a constant unit vector along the line of sight. To evaluate various quantities that appear in the above equation, we apply the Lorentz transformation from the rest frame of $M_p$ to the binary barycenter frame, which leads to 
\begin{equation}
\delta{\vec\beta}={\vec\beta}_{p}-({\vec\beta}\cdot{\vec\beta}_{p}){\vec\beta}+\mathcal{O}(\beta_{p}^2).
\end{equation}
Further, it is straightforward to obtain the derivative of the Doppler factor as 
\begin{equation}
\frac{\partial\ln{\cal D}}{\partial{\vec\beta}}=\frac{\hat{n}}{1-{\hat{n}}\cdot{\vec\beta}}-\frac{\vec\beta}{1-\beta^2}.
\end{equation}
Invoking the fact that for relativistic jets, we have $|{\hat{n}}| = 1$ and $|\vec \beta|=\beta \approx 1$ leads to 
\begin{align}
    ({\hat{n}} - {\vec \beta})\cdot \vec \beta_{p} \sim {\vec \theta}\cdot \vec \beta_{p}.
\end{align}
Note that $\vec \theta$ lies in the plane of the sky, and it leads to
\begin{align}
    {\vec \theta}\cdot{\vec \beta_{p}} \cdot  = {\theta}{\beta_{p}}\cos \iota\cos\left(\frac{2\pi}{P_b} (t-t_0)\right)\,.
\end{align}
It may be worth noting that temporal variation is $\propto \cos \iota $ and it arises because of the scalar product between $\vec \theta$ and $ \vec{\beta}_p$.
It is also straightforward to obtain
\begin{align}
    1-{\hat{n}}\cdot{\vec\beta} = \frac{1+\gamma^2\theta^2}{2\gamma^2}.
\end{align}
Substituting these inputs in Eq.~\ref{eq:fluxchange}  leads to 
\begin{equation}
\delta \ln S=\frac{2(2-\alpha)\gamma^2\theta \beta_{p}\cos \iota }{(1+\gamma^2\theta^2)}\cos\left(\frac{2\pi}{P_b} (t-t_0)\right).
\label{eq:cir1}
\end{equation}
It was argued that the observed amplitude of flux density variations of $0.4$ Jy could be associated with an SMBH binary with total mass $\sim 10^8\, M_{\odot}$ having $ \gamma \sim 10$ \citep{pks_rb}. The $\cos \iota $ dependence requires further explanation, as it is natural to expect that the variations in Doppler factor are prominent when the orbital motion is parallel or anti-parallel to the jet, similar to what we routinely observe in radial velocity measurements \citep{RV_review16}. However, $\delta \ln S $ is $\propto ( \hat{n}- \vec{\beta}) $ and this is $\sim \vec\theta$ which lies in the plane of the sky. This ensures that $( \hat{n}- \vec{\beta} ) \cdot \vec \beta_p $ provides $\cos \iota $ dependence to the leading order $\delta \ln S$ variations. In what follows, we provide a way to incorporate 
orbital eccentricity into the prescription of \cite{pks_rb} for PKS 2131-021.

\subsection{Incorporating the effects of tiny orbital eccentricity into the \cite{pks_rb} description } \label{subsec:ell1}

To incorporate the effects of residual orbital eccentricity into the SMBH binary description for PKS 2131-021 in an efficient manner, we adapt the approach that provided the widely used \texttt{ELL1} timing model \citep{ELL1}. This model is invoked to connect times-of-arrival of pulses from many millisecond pulsars as they reside in tiny eccentricity orbits around their companions \citep{lk2004}. This timing model is critical to improve numerical stability when dealing with binaries that have nearly circular orbits where traditional Keplerian orbital elements become degenerate or meaningless. We begin by briefly listing the Keplerian parametric solution to the Newtonian orbital dynamics of binaries in eccentric orbits which is critical to understand the \texttt{ELL1} timing model. Influenced by \cite{CM_TB}, we describe an eccentric orbit in the polar coordinates and the center of mass reference frame:
\begin{subequations}
\begin{align}
r &= a(1-e\cos u )\,,
\\
\phi -\omega \equiv \nu  &=2 \arctan \biggl [ \biggl ( \frac{ 1 + e}{ 1 - e}
\biggr )^{1/2} \, \tan \frac{u}{2} \biggr ]\,,
\end{align}
\end{subequations}
where $r$ and $\phi$ define the components of the relative separation vector $ \vec{r} = r ( \cos \phi, \sin \phi, 0)$. Further,  the semi-major axis and the eccentricity of the orbit are denoted by $a$ and $e$ respectively, while the auxiliary angles $u$ and $\nu $ are called eccentric and true anomalies respectively. We require to solve the Kepler equation to obtain the temporal evolution of $\vec {r} $ and it reads 
\begin{align}
l \equiv n (t - t_0)  &= u - e\,\sin u\,,
\label{eq:keplereq}
\end{align}
where $l$ is the mean anomaly and $n$ is referred to as the mean motion and is given by $ n = \frac{2\,\pi}{P_b}$, $P_b$ being the orbital period while $t_0$ and $\omega $ provide the epoch of periastron passage and the argument of periastron, respectively. It may be noted that the above description provides the trajectory of the reduced mass and we need to employ appropriate Newtonian relations to connect the center-of-mass (CM) relative motion and CM motion to determine each body's trajectory.

\par
The trajectory of a celestial object moving in three-dimensional space is characterised by its position and velocity vectors. The orbital position as a function of the true anomaly $\nu$ reads 
\begin{subequations}
\begin{align}
    \rm x &= r\cos(\omega + \nu) \\
    \rm y &= r\cos \iota \sin(\omega + \nu) \\
    \rm z &= r\sin \iota \cos(\omega + \nu) \,,
\end{align}
\label{cartesianeccorbit}
\end{subequations}
where orbital inclination $\iota$ is one of the six Keplerian elements and we have 
\begin{align}
    r &= \frac{a(1-e^2)}{1+e\cos\nu} 
\end{align}

\par
To obtain the components of orbital velocity, we differentiate Eqs.~\ref{cartesianeccorbit} and it leads to 
\begin{subequations}
\begin{align}
    \beta_{p\rm x} &= -B [e\sin \omega + \sin (\omega + \nu)] \,,\\
    \beta_{p\rm y} &= B \cos \iota [e\cos \omega + \cos (\omega + \nu)] \,,\\
    \beta_{p\rm z} &= B \sin \iota [e\cos \omega + \cos (\omega + \nu)] \,,
\end{align}  
where 
\begin{equation}
        B = \frac{2\pi a}{c P_b \sqrt{1-e^2}}\,,
\end{equation}
\end{subequations} 
and $\beta_p$ stands for orbital velocity of primary SMBH.

We are now in a position to compute the dot product that appear in the numerator of Eq.~\ref{eq:fluxchange} and we have 
\begin{align}
    (\hat{n} - {\vec \beta})\cdot \vec \beta_{p} \sim {\vec \theta}\cdot \vec \beta_{p} =  \theta_{\rm x} \beta_{p\rm x} + \theta_{\rm y} \beta_{p\rm y} + \theta_{\rm z} \beta_{p\rm z} .
\end{align}

Note that $\theta_{\rm z}$ should be  $0$, as $\vec{\theta}$ lies in the plane of the sky. This leads to 
\begin{align}
     {\vec \theta}\cdot \vec \beta_{p} &=  \theta_{\rm x} \beta_{\rm 1x} + \theta_{\rm y} \beta_{\rm 1y} 
     \nonumber \\
     &=  -B \theta_{\rm x} [e\sin \omega + \sin (\omega + \nu)] +  B \theta_{\rm y} \cos \iota [e\cos \omega + \cos (\omega + \nu)]
     \nonumber\\
     &= B' [e\cos \omega' + \cos (\omega' + \nu)]
     \label{eq:exoeq}
\end{align}
where 
\begin{subequations}
\begin{align}
     B' &= B \sqrt{\theta_{\rm y}^2 \cos^2i + \theta_{\rm x}^2}\\
     \omega' &= \omega + \arcsin \left( \frac{\theta_{\rm x}}{\sqrt{\theta_{\rm y}^2 \cos^2i + \theta_{\rm x}^2}} \right) = \omega + \eta.
\end{align}  
\end{subequations}
The above expression, Eq. \ref{eq:exoeq}, closely resembles the standard radial velocity equation commonly employed in exoplanet detection via the radial velocity method \citep{exoplanets}.

\par
To bring in explicit temporal evolution, we express $\cos\nu$ and $\sin\nu$ in terms of the mean anomaly $l$ while restricting to leading order $e$ contributions
\begin{subequations}
\begin{align}
    \cos \nu &= -e + \frac{2(1-e^2)}{e} \sum_{r = 1}^{\infty} J_r (re)\cos rl \,,
    \nonumber\\
    &= \cos l + (\cos (2l)-1)e + \mathcal{O}(e^2) \,,\\
    \sin \nu &= 2\sqrt{(1-e^2)} \sum_{r = 1}^{\infty} J'_r (re)\sin rl \,,
    \nonumber\\
    &= \sin l + \sin (2l) e + \mathcal{O}(e^2) \,,
\end{align} 
\end{subequations}
where $ J_n(ne)$ stands for Bessel function of the first kind \citep{CM_TB}. This gives us 
\begin{align}
     {\vec \theta}\cdot \vec \beta_{p} 
     &= B' [\cos (\omega' + l) + e \cos (\omega' + 2l)].
\end{align}

Influenced by \cite{ELL1}, we define  $\Phi = \omega' + l$, and it leads to 
\begin{align}
     {\vec \theta}\cdot \vec \beta_{p} 
     &= B' [\cos \Phi + e \cos (2\Phi-\omega')]
     \nonumber\\
     &= B' [\cos \Phi + e \cos \omega' \cos 2 \Phi + e \sin \omega' \sin 2 \Phi].
\label{eq:criticaldotproduct}
\end{align}

We now introduce the Laplace-Lagrange parameters \citep{ELL1} 
\begin{subequations}
\begin{align}
    \epsilon_{\rm 1} &= e \sin \omega'\\
    \epsilon_{\rm 2} &= e \cos \omega'.
\end{align}    
\end{subequations}

This allows us to express 
\begin{subequations}
\label{Eq_22}
\begin{align}
    \Phi &= \omega' + l
    \nonumber\\
    &= n(t-t_{\Omega}')
\end{align}
where 
\begin{align}
    t_{\Omega}' = t_0 - \frac{\omega}{n} - \frac{\eta}{n}.
\end{align}  
\end{subequations}

\par
The critical dot product that appear in Eq.~\ref{eq:fluxchange} for $ \delta \ln S $ becomes 
\begin{align}
{\vec \theta}\cdot \vec \beta_{p} &= B'\left[\cos\Phi + \epsilon_{\rm 2}\cos2\Phi + \epsilon_{\rm 1}\sin2\Phi\right]
\end{align}
where $ \Phi$ is given by $n(t-t_{\Omega}')$. This leads to the following expression for the temporal variations in the radio flux density
\begin{align}
\delta \ln S &= \frac{2(2-\alpha)\gamma^2\theta B'}{(1+\gamma^2\theta^2)}\left[\cos\Phi + \epsilon_{\rm 2}\cos2\Phi + \epsilon_{\rm 1}\sin2\Phi\right]\,,
\label{eq:ell1fluxchange}
\end{align}
where $\Phi$ is given by Eqs.~\ref{Eq_22}, which makes it fully analytic.

\par 
We observe that the temporally evolving part of the above expression is similar to the way  R{\o}mer Delay expression, widely used while timing pulsars in tiny eccentric orbits \citep{ELL1}, changes with time. This delay is due to the light travel time across the pulsar’s orbit because of its motion around the binary centre of mass. This ensures that the pulses arrive earlier when the pulsar is closer and they arrive later when it is  farther with respect to the center of mass and the resulting time variation is known as the binary  R{\o}mer  delay \citep{lk2004}. The fully analytic expression for such a delay for binary pulsars with tiny orbital eccentricities read 
\begin{align}
\Delta_R  & \propto 
				\left\{  
					\sin \Phi
					+\frac{1}{2}\left(
						\epsilon_2 \sin 2\Phi
						-\epsilon_1 ( \cos 2\Phi +3)					
					\right)					
				\right\} + \mathcal{O}(e^2)
\label{romerdelay}
\end{align}
where  $\epsilon_1$ and $\epsilon_2$ are the classical Laplace-Lagrange parameters and $ \Phi  = l+\omega = \frac{2\pi}{P_b}\left(t-t_{\Omega}\right) $ where $t_\Omega$ is a constant \citep{ELL1,ell1k}. It may be noted that this is part of the \texttt{ELL1} timing model, widely used to time millisecond pulsars as they typically exhibit tiny orbital eccentricities \citep{lk2004}. It should be obvious that the temporal evolution of Eq.~\ref{eq:ell1fluxchange} is identical to the time derivative of the above expression. This is not surprising as the SMBH binary model for PKS 2131$-$021, detailed in \cite{pks_rb}, requires modulations in the Doppler boost of the radio jet induced by the orbital motion of $M_s$ around $M_p$. These arguments provide a check for the correctness of our expression for $ \delta \ln S$ given by Eq.~\ref{eq:ell1fluxchange}. Influenced by these discussions, we refer to our eccentric model for PKS~2131-021 as the ELL1 model and explore if the data for PKS~2131-021 employed in \cite{pks_rb} could be used to constrain the values of $\epsilon_1$ and $\epsilon_2$ for this interesting SMBH binary candidate.

\section{ Probing the Presence of Residual Orbital Eccentricity  } \label{sec:obsevid}

We begin by providing a brief description of observational data, detailed in \cite{pks_rb}. 
This is followed by Sec. \ref{sec:obsev}, detailing our Bayesian inference approach to constrain the values of $\epsilon_1$ and $\epsilon_2$ with the help of PKS~2131-021 data. In Sec. \ref{sec:e+drw}, we discuss implications of red noise in the radio lightcurve of PKS~2131-021 and results of Bayesian parameter estimation incorporating DRW red noise process into our analysis.

\subsection{ A Brief Description of the Employed Data} \label{subsec:data}

\begin{figure}[ht!]
\centering
\includegraphics[width=1.0\linewidth]{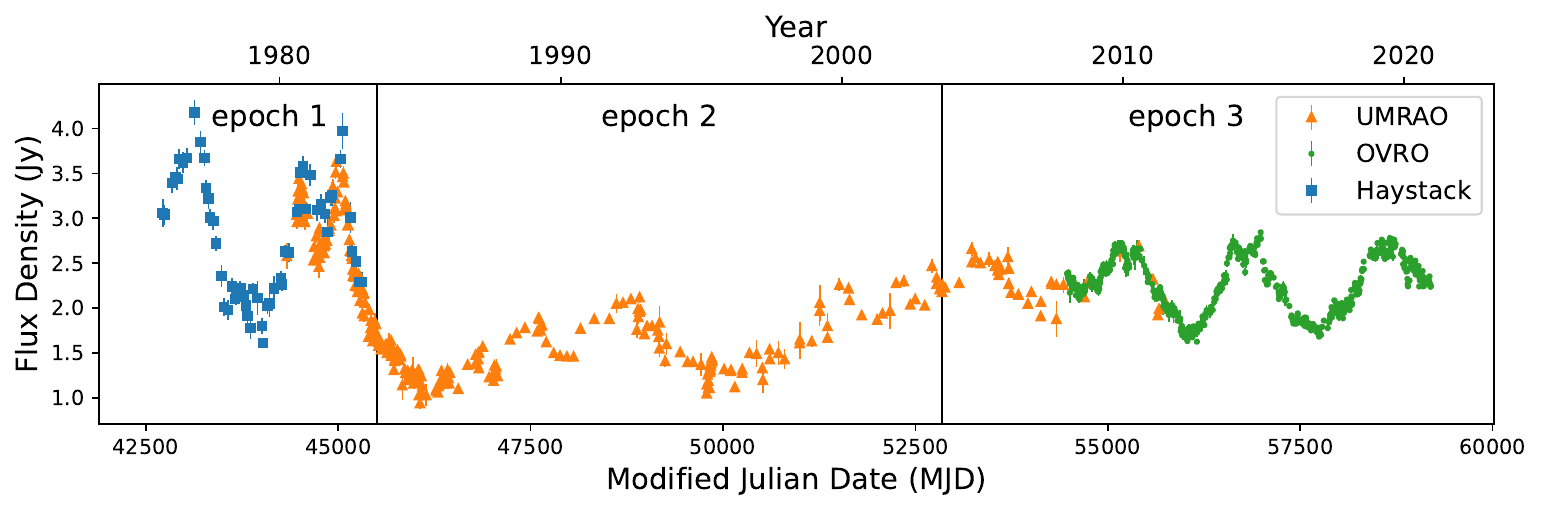}
\caption{Influenced by \citet{pks_rb}, we display the radio light curve of PKS~2131-021 in the 14.5-15.5~GHz frequency range that spans more than four decades. We segregate the data, compiled from observations using Haystack, UMRAO and OVRO, into three temporal intervals and identify sinusoidal flux density variations in  epoch~1 and epoch~3.}
\label{plt:pksdata}
\end{figure}

For the present effort, we employ the radio light curve of PKS 2131-021, a bright quasar at a redshift of $z=1.285$, compiled by \cite{pks_rb} and displayed in Fig.~\ref{plt:pksdata}. The data set of \cite{pks_rb} is compiled from three long-term monitoring programs spanning 45.1 years and it includes observations from the Haystack Observatory (green squares; 15.5 GHz, 1975–1983)~\citep{1986AJ.....92.1262O}, the University of Michigan Radio Astronomy Observatory (UMRAO; brown triangles; 14.5 GHz, 1980–2012)~\citep{1985ApJS...59..513A} , and the Owens Valley Radio Observatory (OVRO; blue dots; 15 GHz, 2008–2021)~\citep{2011ApJS..194...29R}. Further, the median measurement uncertainties are 0.11, 0.05 and 0.03 Jy for Haystack, UMRAO, and OVRO data, respectively. However,  the careful compilation of these data sets by \cite{pks_rb} ensured that the combined light curve offers a nearly continuous view of the flux density variability that spans more than four decades, critical to probe the SMBH binary nature of AGNs as in the case of OJ~287.

\par 
A close inspection of Fig.~\ref{plt:pksdata} reveals three distinct epochs that we separate by vertical black lines. During epoch1 (MJD~$<$~45500), the Haystack and UMRAO data reveal a prominent periodic oscillation, which disappears during epoch2 (45500~$<$~MJD~$<$~52850). However, from the beginning of 2003, the periodic variability reappears with a period and phase remarkably consistent with those seen in epoch1 and this is our  epoch3 (MJD~$>$~52850). Strikingly, both the periodicity and phase between epoch 1 and epoch 3 are consistent within $\sim$2\% and 10\%, respectively. We note that the discovery of a $\sim$4.69-year periodicity in PKS~2131-021 was first reported by \citet{2021MNRAS.506.3791R}, based on 11 years (2008 - 2019) of OVRO monitoring data. This intriguing signature prompted further investigation using archival observations that substantially extended the temporal baseline and allowed a more robust assessment of the periodic behaviour \citep{pks_rb}. The persistence and reappearance of this periodic behaviour across decades make PKS~2131$-$021 a compelling candidate for an SMBH binary system.

\subsection{Observational Evidence for the Presence of Tiny Orbital Eccentricity }\label{sec:obsev}

We perform Bayesian inference and model selection as described in Appendix~\ref{sec:method} to measure the eccentricity of this system. Guided by \cite{pks_rb} and our Eq.~\ref{eq:cir1}, we express the signal 
$s\ (= S_0 + \delta S)$ for the circular model as:
\begin{equation}
   {s}(t)=S_{0} + A\cos\left(\frac{2\pi}{P_b}(t-t_{0})\right)\,.
\label{eq:signaKO}
\end{equation}
In contrast, the expected signal from our eccentric ELL1 model, influenced by Eq. \ref{eq:ell1fluxchange}, reads
\begin{equation}
   {s}(t)=S_{0} + A\left[\cos\left(\frac{2\pi} {P_b} (t - t_{\Omega}')\right) + \epsilon_{\rm 1}\cos2\left(\frac{2\pi} {P_b} (t - t_{\Omega}')\right) + \epsilon_{\rm 2}\sin2\left(\frac{2\pi} {P_b} (t - t_{\Omega}')\right)\right]\,,
\label{eq:signaeKO}
\end{equation}
where $S_{0}$ stands for the mean flux density from both core and jet i.e. $S_{0} = S_{core} + S_{jet}$ and $A$ is the amplitude of the signal that depends on parameters $S_{jet}, \alpha, B, \iota, \gamma, $ and $\theta$. The model parameters for circular and eccentric ELL1 models are  $   \Theta=\{S_{0},A,t_{0},P_b,\sigma_{0}\}$ and $   \Theta=\{S_{0},A,t_{\Omega}',P_b,\epsilon_{1},\epsilon_{2},\sigma_{0}\}$, respectively and it should be obvious that our ELL1 model introduces two additional parameters. Here, $P_b$ represents the orbital period in the observer's frame, from which the rest-frame period of the SMBHB can be derived as $P_{\rm rest} = P_b/(1+z)$.

\par
To pursue Bayesian parameter estimation(PE) studies described in Appendix~\ref{sec:method}, we employed the \texttt{Nautilus} Python package~\citep{nautilus} for Bayesian posterior and evidence estimation. This computationally efficient package invokes neural networks that require fewer likelihood calls and this ensures that the approach is faster than the traditional Markov Chain Monte Carlo (MCMC) and nested sampling algorithms. To assure reliability, we have independently repeated the sampling procedure several times. The resulting values of ln $\mathcal{Z}$ remained consistent within the expected uncertainties, and further, the posterior distributions demonstrated stable convergence throughout the runs.

\par 
For the present efforts, we employed four datasets: OVRO (O), OVRO + Haystack (OH), OVRO + UMRAO (OU) and OVRO + Haystack + UMRAO (OHU). In all of these data sets, we omitted epoch 2 data from UMRAO and considered only epoch 1 and epoch 3 data for the analysis, which contains periodic variations as done in \cite{pks_rb}. Further, we used different $S_0$ and $A$ for epoch1 and epoch3, as they show different mean offset and amplitude. We also used different EQUAD parameters for different observatories. The remaining parameters $P_b$, $\phi_0$, $\phi_\Omega'$, $\epsilon_1$ and $\epsilon_2$ are the same for all the observatory data and all the epochs. The priors used in our Bayesian PE runs are displayed in Table \ref{tab:prior_details} and we restrict the priors on $\epsilon_1$ and $\epsilon_2$ within a tiny range of $[-0.1,0.1]$. This is mainly because  Eq. \ref{eq:signaeKO} for our ELL1 model is only valid for tiny eccentricities \textbf{i.e. $e < 0.12$ as discussed in Appendix ~\ref{app:uncertainty}} and astrophysical considerations support the possibility that such SMBH binaries may have residual eccentricities \citep{astro_e}.

\par 
We now display with the help of three tables various results of our Bayesian PE studies. Posterior medians and $1\sigma$ credible intervals of all model parameters for both circular and ELL1 SMBH binary descriptions are tabulated in Table \ref{tab:posterior_circular} and Table \ref{tab:posterior_ELL1}, respectively while employing different datasets. The corresponding natural logarithm of evidence ($\ln \mathcal{Z}$) and associated Bayes factors ($\ln \mathcal{B}^{ELL1}_{cir}$) are listed in Table. \ref{tab:bayes_transposed}. These entries suggest strong data based evidence for the presence of residual eccentricity in the SMBH binary prescription for PKS~2131-021.

\par
\begin{figure}[htbp]
    \centering
    \begin{subfigure}[b]{0.45\textwidth}
        \includegraphics[width=\textwidth]{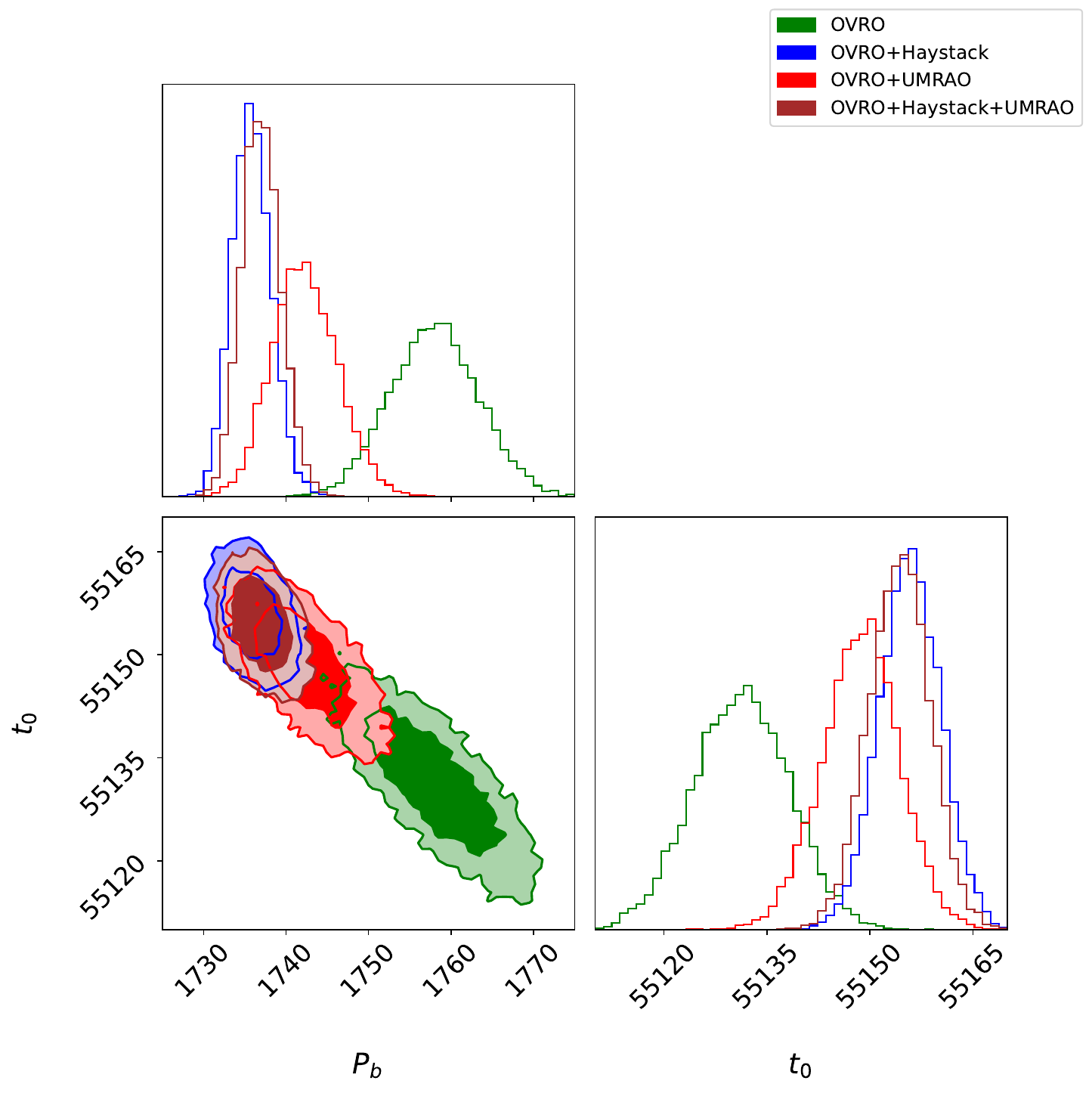}

        \label{fig:cir_comb}
    \end{subfigure}
    \hfill
    \begin{subfigure}[b]{0.45\textwidth}
        \includegraphics[width=\textwidth]{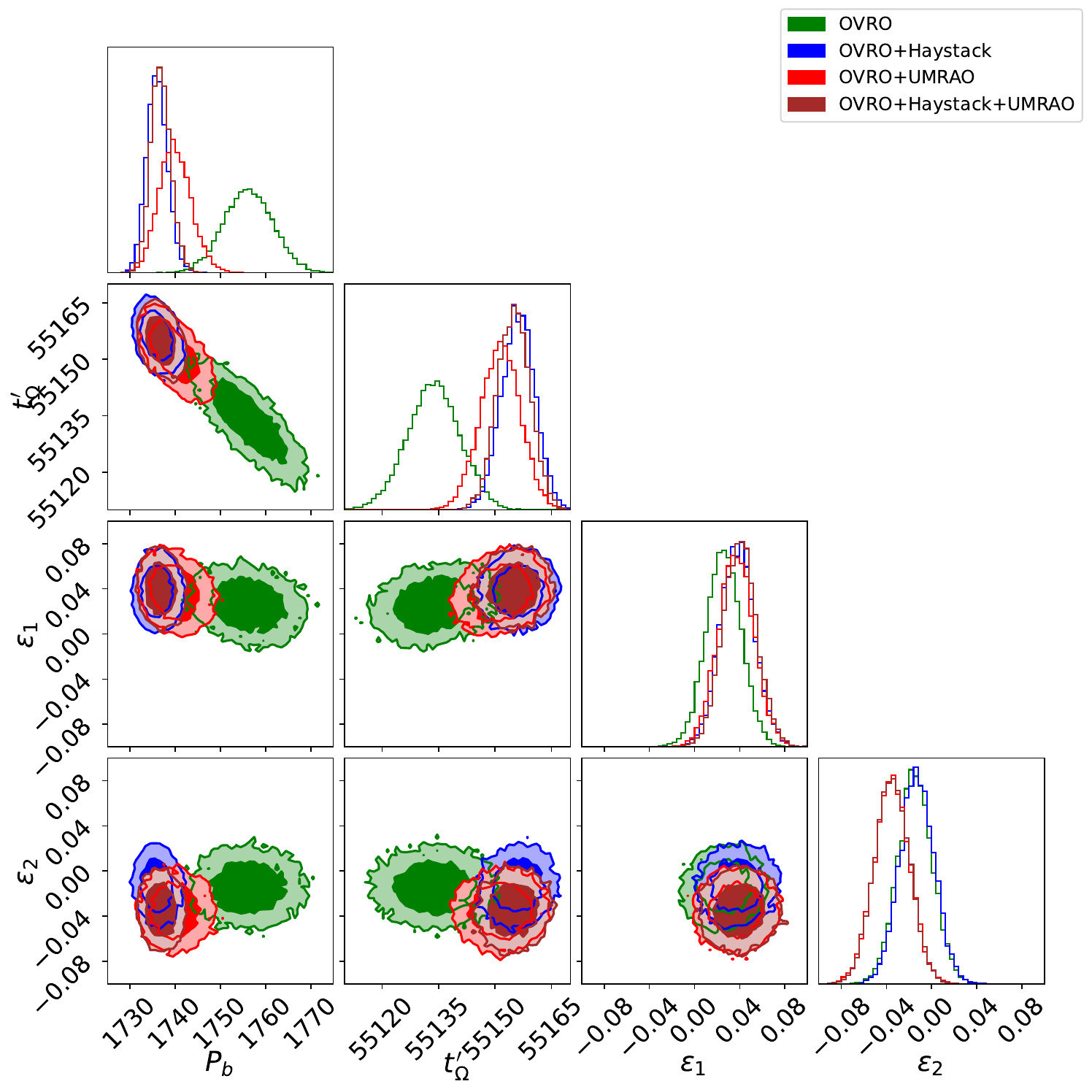}

        \label{fig:ell1_comb}
    \end{subfigure}
    \caption{Summary/comparative plots that provide posterior distributions associated with our circular (left) and  ELL1 (right) models while employing various combinations of available OVRO, Haystack, and UMRAO data sets. Interestingly, constraints on the LL-like parameters are consistent across various data sets. }
    \label{fig:comb}
\end{figure}

 
We now display posterior distributions of important binary parameters that arise from our PE runs while employing combinations of various data sets and the two theoretical prescriptions. In other words, we show posterior distributions of binary parameters $\{P_b, t_0\}$ for the circular model and relevant parameters $\{P_b, t_\Omega', \epsilon_1, \epsilon_2\}$ for our ELL1 binary model in Fig. \ref{fig:comb}. We now itemize the key results from these plots for easy reading:
\begin{itemize}
    \item 
For the OVRO(O) dataset, the circular model leads to a binary with an orbital period $1758\pm5$ days and our ELL1 model indicates the presence of eccentric binary with an orbital period $1756\pm5$ days. From the posteriors of LL-like parameters, we infer the orbital eccentricity to be $0.030\pm0.015$ while $\omega' = 120^\circ\pm30^\circ$. Interestingly, the natural logarithm of the Bayes factor i.e. $\ln \mathcal{B}^{\rm ELL1}_{\rm cir} = -1.40$ implies a rather negative support for the ELL1 model.
    \item
For the combined OVRO+Haystack (OH) data, the circular model, as expected, supports the presence of circular binary with a period of $1736\pm2$ days. In contrast, the ELL1 also supports a binary with an orbital period  of $1736\pm2$ days while eccentricity  and $\omega'$ parameters take values $0.040\pm0.015$ and $109^\circ\pm22^\circ$, respectively . The natural logarithm of the Bayes factor $\ln \mathcal{B}^{\rm ELL1}_{\rm cir} = +0.26$ and this implies support worth no more than bare mention for the ELL1 model.
    \item
For another combined data sets, namely OVRO+UMRAO (OU) data set, the circular model provides a binary with an orbital period of $1742\pm4$ days while our ELL1 model indicates the presence of eccentric binary with a period of $1740\pm4$ days. The LL-like parameter posteriors lead to $e =0.051\pm0.016$ and $\omega'= 133^\circ\pm18^\circ$. Interestingly, the natural logarithm of the Bayes factor $\ln \mathcal{B}^{\rm ELL1}_{\rm cir} = +2.22$ which lends positive support for the ELL1 model.
    \item 
Finally, the full  OVRO+Haystack+UMRAO (OHU) data set supports the circular model with an orbital period of  $1737\pm2$ days. In contrast, the ELL1 model indicates the presence of a tiny eccentric SMBH binary with a period of $1737\pm2$ days while the LL-like parameter posteriors provide $ e= 0.053\pm0.015$ and $\omega'= 131^\circ\pm16^\circ$. Further, natural logarithm of the Bayes factor, namely  $\ln \mathcal{B}^{\rm ELL1}_{\rm cir} = +3.15$, implies a rather strong support for the ELL1 model.
\end{itemize}

\par 
The Fig. \ref{fig:comb} summarizes all these binary parameter posteriors to show how these important binary posteriors for various data sets and two models overlap with one another. We infer that posterior distributions of the orbital period becomes narrower with the addition of more data which essentially increases the time baseline. Further, the posteriors of  $P_b$, $t_0$ and $t_\Omega'$ parameters for datasets OH, OU and OHU are consistent within $1\sigma$ though there is some tension while dealing only with the OVRO data, possibly due to a smaller time baseline. Interestingly, the posteriors of $\epsilon_1$ and $\epsilon_2$ are fairly consistent with each other for all the datasets. We note from the entries in the Table. \ref{tab:posterior_circular} and Table. \ref{tab:posterior_ELL1} that other common parameters, including the reference epochs $t_0$ and $t_\Omega'$ associated with the two models, are consistent between circular and ELL1  models for the considered datasets. These considerations, along with the inferred significant deviation of $\epsilon_1$ and $\epsilon_2$ from $0$, as required from a circular binary description, strongly support the presence of residual eccentricity in these data sets. Additionally, we observe an increasing support for our eccentric ELL1 model with the inclusion of additional data that also increases the lightcurve time baseline, as evident from the $\ln \mathcal{B}^{\rm ELL1}_{\rm cir}$ entries tabulated in Table. \ref{tab:bayes_transposed}. We note while passing that the other parameters like ${\rm S0\_epoch3}$ and ${\rm A\_epoch3}$ are stable over different datasets. In contrast, ${\rm S0\_epoch1}$ and ${\rm A\_epoch1}$ vary significantly for different datasets, which can be primarily attributed to fewer data points and smaller time baseline in the Haystack and UMRAO data used for the estimation of these parameters. We also note that the EQUAD parameters for OVRO, Haystack and UMRAO are also consistent among analysis of different datasets.

\begin{deluxetable*}{lllll}
\tablecaption{Prior Distributions used in our Bayesian PE efforts \label{tab:prior_details}}
\tablehead{
\colhead{Parameter} & \colhead{Description} & \colhead{Prior Type} & \colhead{Range} & \colhead{Units}
}
\startdata
S0\_epoch1     & Flux density at epoch 1           & Uniform & [$1.5, 4.5$]        & Jy     \\
S0\_epoch3     & Flux density at epoch 3           & Uniform & [$1.5, 3.0$]        & Jy     \\
A\_epoch1      & Amplitude at epoch 1              & Uniform & [$0, 1.5$]          & Jy     \\
A\_epoch3      & Amplitude at epoch 3              & Uniform & [$0, 0.75$]         & Jy     \\
$P_b$            & Orbital period                    & Uniform & [$1500, 2000$]      & days   \\
$t_0$          & Reference time        & Uniform & [$54000, 55700$]    & MJD    \\
EQUAD\_OVRO    & Additional white noise (OVRO)     & Uniform & [$0, 5$]            & Jy    \\
EQUAD\_UMRAO   & Additional white noise (UMRAO)    & Uniform & [$0, 5$]            & Jy    \\
EQUAD\_Haystack& Additional white noise (Haystack) & Uniform & [$0, 5$]            & Jy    \\
$\epsilon_1$          & Eccentricity component ($e \sin\omega$) & Uniform & [--0.1, 0.1]     & ---    \\
$\epsilon_2$          & Eccentricity component ($e \cos\omega$) & Uniform & [--0.1, 0.1]     & ---    \\
$\ln \hat{\sigma}$    & Intrinsic variance     & Uniform & [$-6, 0$]            & Jy day$^{0.5}$   \\
$\ln \tau_{0}$   & Damping timescale    & Uniform & [$-4, 15$]            & days    \\
$\gamma$ & exponent & Uniform & [$0, 1.8$]            &  ---   \\
$\log_{10}e$          & Eccentricity & Uniform & [$-6, 0$]     & ---    \\
$\omega$          & Argument of Periastron & Uniform & [$-\pi, \pi$]     & radians    \\
\enddata
\tablecomments{All priors are uniform across the indicated ranges and  
units are absent for dimensionless parameters.}
\end{deluxetable*}

\begin{deluxetable*}{lcccc}
\tablecaption{Posterior estimates for circular SMBH binary model for different datasets. \label{tab:posterior_circular}}
\tablehead{
\colhead{Parameter} & \colhead{O} & \colhead{OH} & \colhead{OU} & \colhead{OHU}
}
\startdata
S0\_epoch1 (Jy)      & ---                                & $2.70^{+0.04}_{-0.04}$   & $2.56^{+0.04}_{-0.04}$    & $2.57^{+0.03}_{-0.03}$    \\
S0\_epoch3 (Jy)      & $2.225^{+0.005}_{-0.005}$          & $2.229^{+0.005}_{-0.004}$ & $2.228^{+0.005}_{-0.005}$ & $2.229^{+0.005}_{-0.005}$ \\
A\_epoch1 (Jy)       & ---                                & $0.84^{+0.06}_{-0.06}$    & $0.56^{+0.06}_{-0.06}$    & $0.68^{+0.05}_{-0.05}$    \\
A\_epoch3 (Jy)       & $0.408^{+0.007}_{-0.007}$          & $0.409^{+0.007}_{-0.007}$ & $0.408^{+0.007}_{-0.007}$ & $0.408^{+0.007}_{-0.007}$ \\
$P_b$ (days)          & $1758^{+5}_{-5}$                  & $1736^{+2}_{-2}$           & $1742^{+4}_{-4}$           & $1737^{+2}_{-2}$           \\
$t_0$ (MJD)          & $55131^{+7}_{-7}$                 & $55156^{+4}_{-4}$           & $55149^{+5}_{-6}$           & $55154^{+4}_{-5}$           \\
EQUAD\_OVRO          & $0.114^{+0.003}_{-0.003}$         & $0.116^{+0.004}_{-0.003}$   & $0.115^{+0.003}_{-0.003}$   & $0.116^{+0.003}_{-0.003}$   \\
EQUAD\_UMRAO         & ---                                & ---                        & $0.30^{+0.02}_{-0.02}$     & $0.31^{+0.02}_{-0.02}$     \\
EQUAD\_Haystack      & ---                                & $0.26^{+0.03}_{-0.03}$     & ---                        & $0.32^{+0.04}_{-0.04}$     \\
\enddata
\tablecomments{Posterior medians and $1\sigma$ credible intervals for circular SMBHB model are reported. Note that the dashes (---) indicate that these parameters are not required for the corresponding dataset.}
\end{deluxetable*}

\begin{deluxetable*}{lcccc}
\tablecaption{Posterior estimates of our eccentric ELL1 SMBH binary model for different datasets.\label{tab:posterior_ELL1}}
\tablehead{
\colhead{Parameter} & \colhead{O} & \colhead{OH} & \colhead{OU} & \colhead{OHU}
}
\startdata
S0\_epoch1 (Jy)      & ---                                & $2.70^{+0.04}_{-0.04}$    & $2.54^{+0.04}_{-0.04}$   & $2.57^{+0.03}_{-0.03}$  \\
S0\_epoch3 (Jy)      & $2.225^{+0.005}_{-0.005}$          & $2.23^{+0.005}_{-0.004}$  & $2.229^{+0.004}_{-0.005}$& $2.229^{+0.004}_{-0.005}$\\
A\_epoch1 (Jy)       & ---                                & $0.84^{+0.05}_{-0.05}$    & $0.59^{+0.06}_{-0.06}$   & $0.69^{+0.04}_{-0.04}$  \\
A\_epoch3 (Jy)       & $0.409^{+0.006}_{-0.006}$          & $0.409^{+0.006}_{-0.006}$ & $0.409^{+0.007}_{-0.006}$& $0.409^{+0.006}_{-0.006}$\\
$P_b$ (days)         & $1756^{+5}_{-5}$                   & $1736^{+2}_{-2}$          & $1740^{+4}_{-3}$         & $1737^{+2}_{-2}$        \\
$t_\Omega'$ (MJD)   & $55134^{+7}_{-7}$                  & $55156^{+4}_{-4}$         & $55151^{+5}_{-5}$        & $55155^{+4}_{-4}$       \\
EQUAD\_OVRO          & $0.114^{+0.003}_{-0.003}$          & $0.116^{+0.003}_{-0.003}$ & $0.115^{+0.003}_{-0.003}$& $0.116^{+0.003}_{-0.003}$\\
EQUAD\_UMRAO         & ---                                & ---                       & $0.29^{+0.02}_{-0.02}$   & $0.3^{+0.02}_{-0.02}$   \\
EQUAD\_Haystack      & ---                                & $0.26^{+0.03}_{-0.03}$    & ---                      & $0.32^{+0.04}_{-0.04}$  \\
$\epsilon_1$   & $0.026^{+0.015}_{-0.015}$ & $0.038^{+0.015}_{-0.015}$ & $0.037^{+0.016}_{-0.016}$ & $0.040^{+0.015}_{-0.015}$ \\
$\epsilon_2$   & $-0.015^{+0.016}_{-0.015}$ & $-0.013^{+0.015}_{-0.015}$ & $-0.035^{+0.015}_{-0.016}$ & $-0.035^{+0.015}_{-0.015}$ \\
\enddata
\tablecomments{Posterior medians and $1\sigma$ credible intervals for ELL1 SMBHB model are reported. Note that the dashes (---) indicate that these parameters are not required for the corresponding dataset.}
\end{deluxetable*}

\begin{deluxetable*}{lcccc}
\tablecaption{Bayesian Evidence and Bayes Factors for Circular vs.\ ELL1 Models\label{tab:bayes_transposed}}
\tablehead{
\colhead{Quantity} & \colhead{O} & \colhead{OH} & \colhead{OU} & \colhead{OHU}
}
\startdata
$\ln Z$ (Circular)               & 489.27 & 462.04 & 442.45 & 413.46 \\
$\ln Z$ (ELL1)                   & 487.87 & 462.30 & 444.67 & 416.61 \\
$\ln \mathcal{B}^{\rm ELL1}_{\rm cir}$ & $-1.40$ & $+0.26$ & $+2.22$ & $+3.15$ \\
\enddata
\tablecomments{Comparison of natural log evidences ($\ln Z$) and natural log Bayes factors for the 
earlier circular and our eccentric ELL1 models across different data configurations.
Note that positive $\ln \mathcal{B}$ values suggest a preference for the ELL1 model.}
\end{deluxetable*}

\subsection{Search for Orbital Eccentricity in the presence of red noise}\label{sec:e+drw}

It is well known that quasars exhibit stochastic variability in their light curves, and such red noise processes can mimic periodic signals, leading to false positive detections of SMBH binary signatures \citep{Vaughan2016, Witt2022, obssmbhb}. To investigate the effect of red noise in the radio light curve of PKS 2131‑021 during our search for residual eccentricity, we performed simulation studies.
To begin, we simulated a sinusoidal signal (i.e., an SMBHB signature without eccentricity) along with artificial red noise derived from the noise properties of the OVRO data using \texttt{lcsim}\footnote{\url{https://github.com/skiehl/lcsim}} \citep{lcsim}. We then analyzed these artificial simulated radio light curves with our ELL1 SMBHB model given in Eq.~\ref{eq:signaeKO}. Our analysis supported non-zero eccentricity, even though the artificially generated light curves contained SMBHB signatures with zero eccentricity. This leads us to conclude that red noise present in quasar light curves can mimic a small eccentricity, and that stochastic red noise processes should be properly accounted for during the analysis.

Prompted by our numerical simulations, we now reanalyze the data using a Damped Random Walk (DRW) red-noise process alongside the eccentric signal within our Bayesian analysis framework. The DRW red noise model is one of the most successful models for the stochastic variability in quasar light curves \citep{obssmbhb, McLoed2010, Kozlowski2010} and is suitable for describing the red noise process in the current data. Because the introduction of DRW red noise into our analysis requires much broader priors on the eccentricity, the previous ELL1 SMBHB model is no longer valid. We therefore use the full eccentricity expression given by Eq.~\ref{eq:exoeq} as:
\begin{equation}
   {s}(t)=S_{0} + A\left[e\cos \omega' + \cos (\omega' + \nu)\right].
\label{eq:fulleKO}
\end{equation}
We use \texttt{kepler.py} \footnote{\url{https://github.com/dfm/kepler.py}} python package to evaluate the $\cos \nu$ and $\sin \nu$ functions, required in the above equation.
Thereafter, we add the covariance matrix $\mathrm{C}_\mathrm{red}$, representing DRW red noise, to the white noise covariance matrix $\mathrm{C}_\mathrm{white}$ described in Eq.~\ref{eq:Cwhite}. The combined noise matrix is then given by $\mathrm{C}_{ij} = \mathrm{C}_\mathrm{red} + \mathrm{C}_\mathrm{white}$, where $\mathrm{C}_\mathrm{red}$ is expressed as:
\begin{equation}
\mathrm{C}_{ij, \mathrm{red}} = \frac{1}{2} \hat\sigma^{2} \tau_{0} \exp \left[-\left(\frac{\tau_{ij}}{\tau_{0}}\right)^\gamma\right],
\label{eq:Cwhite}
\end{equation}
where  $\hat\sigma^2$ stands for the intrinsic variation between observations on short timescales, $\tau_{0}$ is the damping timescale, and $\tau_{ij} = |t_i - t_j|$.
Further, we introduce 
$\gamma$  to account for the possibility of deviation from DRW red noise, and $\gamma = 1$ corresponds to the DRW model \citep{zhusuperbayes, Kelly2009, Guo2017, Zu2013}. 
This ensures that the model parameters for this eccentric SMBHB model are $   \Theta=\{S_{0},A,t_{\Omega}',P_b,\epsilon_{1},e,\omega'\}$.

The priors used for our Bayesian analysis with the DRW model are also displayed in Table \ref{tab:prior_details}, and we use log uniform priors on eccentricity in the range $[-6,0]$ as we expect tiny eccentricity for the PKS 2131-021 SMBHB candidate. We also perform the Bayesian analysis with DRW + Circular SMBHB model for the model comparison. We display the results from our Bayesian analysis for Circular SMBHB + DRW and Eccentric SMBHB + DRW red noise model tabulated in Table ~\ref{tab:posterior_circular_DRW} and \ref{tab:posterior_eccDRW}, similar to results presented in Sec.~\ref{sec:obsev}. The corresponding natural logarithm of evidence ($\ln \mathcal{Z}$) and associated Bayes factors ($\ln \mathcal{B}^\mathrm{ecc+DRW}_\mathrm{cir+DRW}$) are listed in Table. \ref{tab:bayes_transposed_DRW}. These entries suggest no evidence for the presence of residual eccentricity in the SMBH binary prescription for PKS~2131-021.

\par
\begin{figure}[ht!]
\centering
\includegraphics[width=1.0\linewidth]{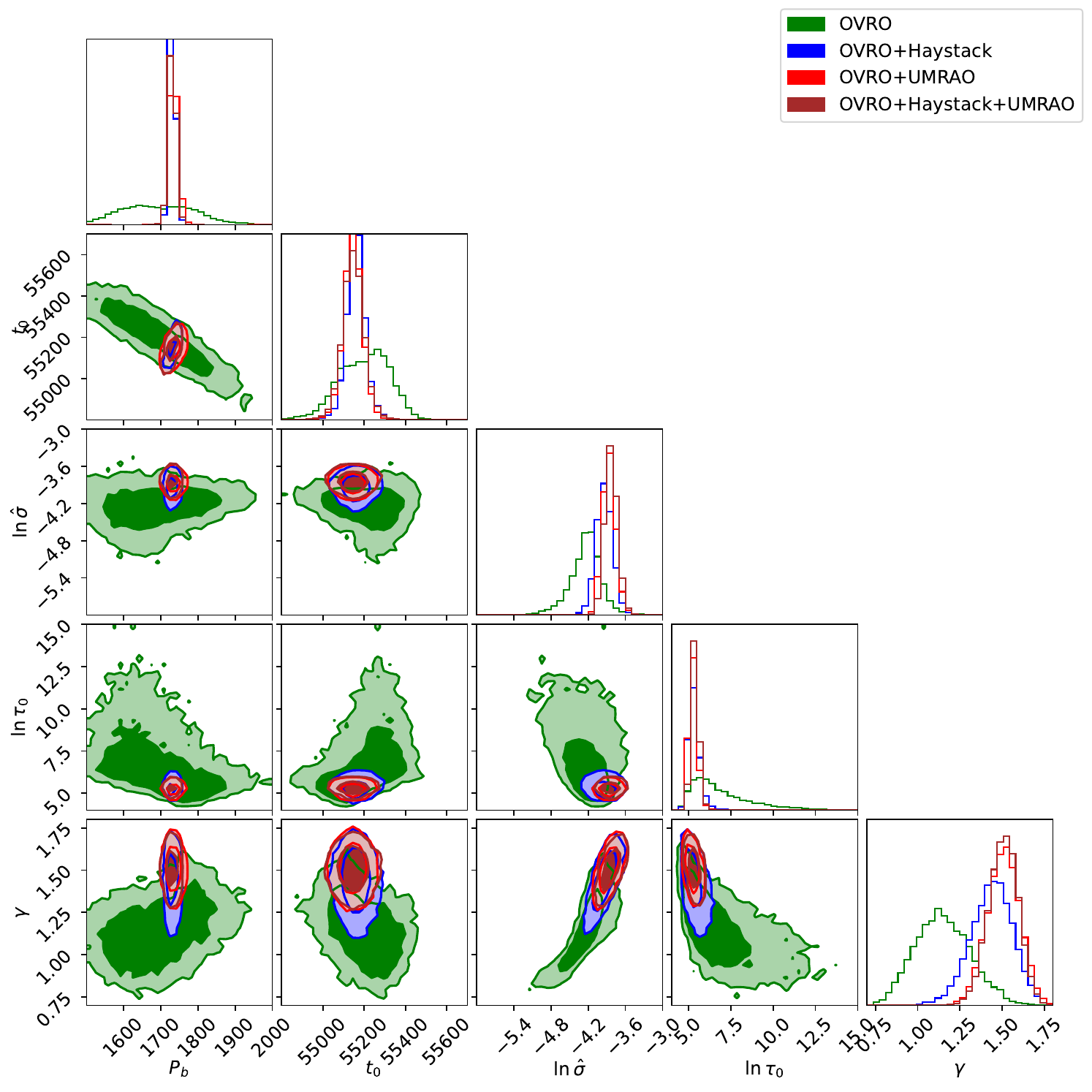}
\caption{Summary/comparative plots that provide posterior distributions associated with our circular SMBHB + DRW rednoise models while employing various combinations of available OVRO, Haystack, and UMRAO data sets.}
\label{fig:cirDRW}
\end{figure}

\begin{figure}[ht!]
\centering
\includegraphics[width=1.0\linewidth]{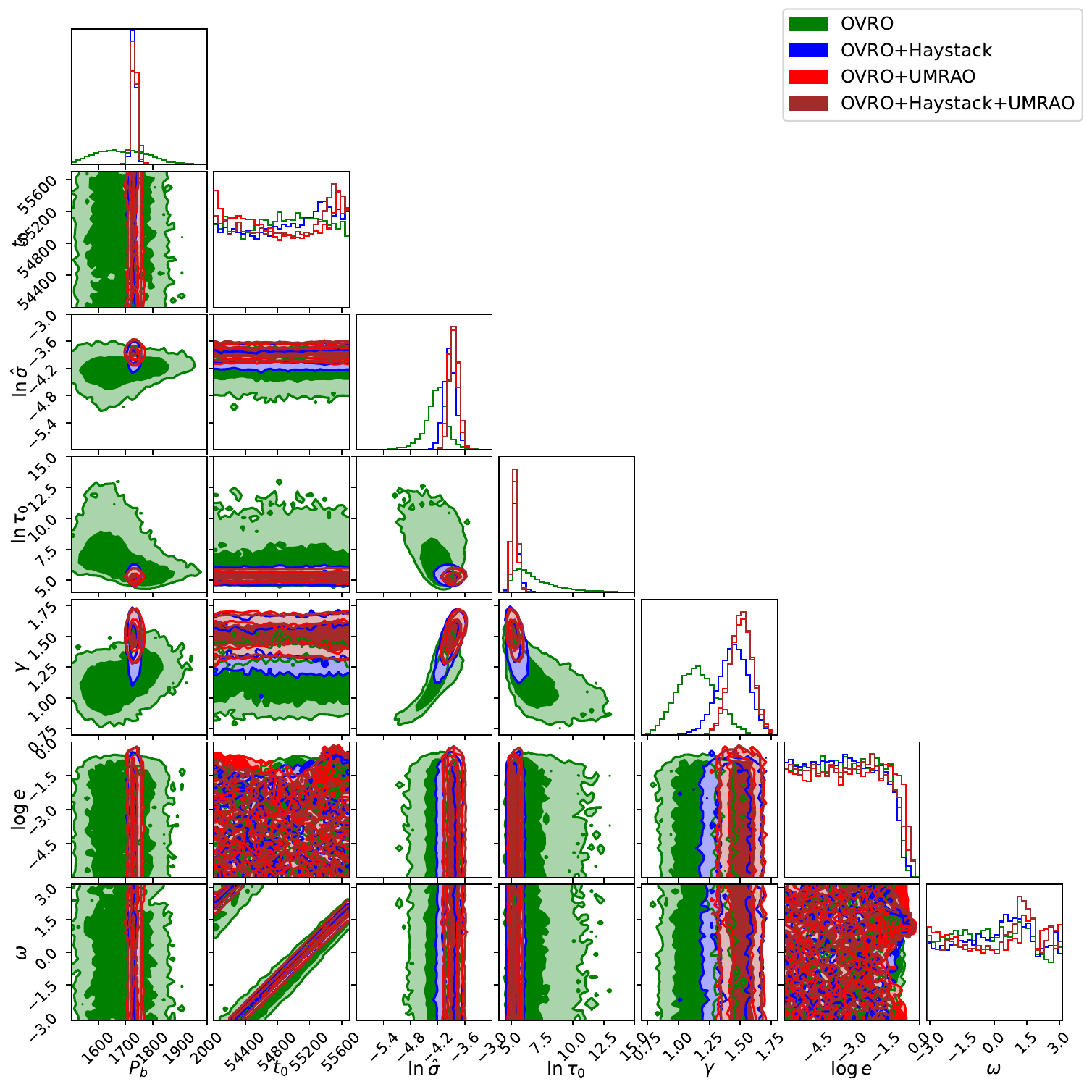}
\caption{Summary/comparative plots that provide posterior distributions associated with our eccentric SMBHB + DRW rednoise models while employing various combinations of available OVRO, Haystack, and UMRAO data sets. }
\label{fig:logkDRW}
\end{figure}

 
The posterior distributions of important parameters, including the DRW red noise parameters from our Bayesian runs, are shown in Figs. \ref{fig:cirDRW} and \ref{fig:logkDRW} for Circular + DRW and Eccentric + DRW models respectively for the four datasets mentioned in Sec.~\ref{sec:obsev}. We can clearly observe that the parameters of our current red noise model are well constrained and agree with each other in various datasets. We can also note that parameters $t_0$ and $\omega'$ are not constrained in the case of the Eccentric + DRW model, due to high covariance between those parameters. In all the datasets, eccentricity is not constrained, and we can only place an upper limit. We now proceed to itemize the key results for different datasets:
\begin{itemize}
    \item 
For the OVRO(O) dataset, the circular + DRW model leads to a binary with an orbital period $1693\pm103$ days, and our Eccentric + DRW model indicates the presence of an eccentric binary with an orbital period $1684\pm104$ days. We have large uncertainties in the orbital period because of an additional red noise process and a smaller data span in this analysis. From the posteriors of $\log_{10}e$, we infer the $95\%$ upper limit on the orbital eccentricity to be $\log_{10}e < -1.07$ or $e < 0.08$. Interestingly, the natural logarithm of the Bayes factor is $\ln \mathcal{B}^{\rm ecc+DRW}_{\rm cir+DRW} = -0.16$, which indicates no support for the eccentric + DRW model.
    \item
For the combined OVRO+Haystack (OH) data, the circular + DRW model, as expected, supports the presence of a circular binary with a period of $1731\pm8$ days. In contrast, the Eccentric + DRW model also supports a binary with an orbital period of $1731\pm8$ days, while a $95\%$ upper limit on the eccentricity is $\log_{10}e < -1.14$ or $e < 0.07$. The natural logarithm of the Bayes factor $\ln \mathcal{B}^{\rm ecc+DRW}_{\rm cir+DRW} = -0.20$ and this implies no support for the eccentric + DRW model.
    \item
For another combined data set, namely OVRO+UMRAO (OU) data set, the circular + DRW model provides a binary with an orbital period of $1734\pm12$ days, while our Eccentric + DRW indicates the presence of a binary with a period of $1734\pm12$ days. The eccentricity parameter posteriors lead to a $95\%$ upper limit $\log_{10}e < -0.81$ or $e < 0.15$. The natural logarithm of the Bayes factor $\ln \mathcal{B}^{\rm ecc+DRW}_{\rm cir+DRW} = -0.13$, which also provides no support for the eccentric + DRW model.
    \item 
Finally, the full  OVRO+Haystack+UMRAO (OHU) data set supports the circular + DRW model with an orbital period of  $1731\pm9$ days. In contrast, the Eccentric + DRW indicates the presence of a SMBH binary with a period of $1731\pm9$ days, while the eccentricity parameter posteriors provide a $95\%$ upper limit $\log_{10}e < -0.86$ or $e < 0.13$. Further, the natural logarithm of the Bayes factor, namely  $\ln \mathcal{B}^{\rm ecc+DRW}_{\rm cir+DRW} = -0.15$, and this implies no support for the eccentric + DRW model.
\end{itemize}


We infer from the entries in Tables~\ref{tab:bayes_transposed} and~\ref{tab:bayes_transposed_DRW} that $\ln \mathcal{B}^{\rm cir+DRW}_{\rm cir} > 600$ for all four datasets. This strongly indicates that the Circular+DRW model is overwhelmingly favored over the Circular model pursued in previous works. In other words, our results show that the preferred description of PKS 2131‑021 is an SMBH binary in a circular orbit, whose signal is embedded in strong, stochastic red noise. Additionally, we observe that the signal parameter uncertainties increase substantially when we incorporate the DRW process, compared with earlier efforts that ignored red noise in the data \citep{pks_rb}.
In particular, the orbital period uncertainty grows significantly when moving from the pure circular model to the Circular+DRW model across all four datasets. This increase in period uncertainty likely led to the disappearance of support
for a tiny orbital eccentricity, while still allowing the presence of red noise in the radio light curve. Possible reasons include the fact that eccentricity and the argument of periastron encode phase-sensitive deviations from a sinusoid at higher harmonics; a poorly constrained period naturally limits our ability to distinguish a small eccentricity signal from noise fluctuations. It is therefore plausible that the absence of evidence for eccentricity in the DRW runs is a consequence of larger error bars on the binary period.

Nevertheless, the present effort uncovers an interesting fact: the Circular+DRW model continues to recover a coherent periodic signal across all datasets, with the orbital period remaining well-defined despite the broader uncertainties. This persistence of periodicity, even in the presence of an explicit stochastic red noise component, provides further support for the underlying SMBH binary interpretation of PKS 2131‑021. It strengthens the conclusion that the observed sinusoidal modulation is not merely an artifact of blazar variability.

\begin{deluxetable*}{lcccc}
\tablecaption{Posterior estimates for circular SMBH binary + DRW red noise model for different datasets. \label{tab:posterior_circular_DRW}}
\tablehead{
\colhead{Parameter} & \colhead{O} & \colhead{OH} & \colhead{OU} & \colhead{OHU}
}
\startdata
S0\_epoch1	&	---	&	$2.68^{+0.07}_{-0.08}$	&	$2.48^{+0.11}_{-0.11}$	&	$2.60^{+0.08}_{-0.08}$	\\
A\_epoch1	&	---	&	$0.88^{+0.10}_{-0.10}$	&	$0.64^{+0.14}_{-0.14}$	&	$0.80^{+0.11}_{-0.11}$	\\
EQUAD\_UMRAO	&	---	&	---	&	$0.10^{+0.01}_{-0.01}$	&	$0.097^{+0.01}_{-0.01}$	\\
EQUAD\_Haystack	&	---	&	$0.11^{+0.03}_{-0.03}$	&	---	&	$0.17^{+0.03}_{-0.03}$	\\
S0\_epoch3	&	$2.27^{+0.22}_{-0.11}$	&	$2.25^{+0.05}_{-0.05}$	&	$2.25^{+0.05}_{-0.05}$	&	$2.25^{+0.05}_{-0.05}$	\\
A\_epoch3	&	$0.35^{+0.06}_{-0.07}$	&	$0.36^{+0.07}_{-0.07}$	&	$0.34^{+0.06}_{-0.06}$	&	$0.33^{+0.07}_{-0.07}$	\\
$P_b$	&	$1693^{+103}_{-98}$	&	$1731^{+8}_{-7}$	&	$1734^{+12}_{-11}$	&	$1731^{+9}_{-9}$	\\
$t_0$	&	$55208^{+112}_{-132}$	&	$55159^{+43}_{-41}$	&	$55145^{+44}_{-44}$	&	$55147^{+50}_{-50}$	\\
$\ln\hat{\sigma}$	&	$-4.22^{+0.20}_{-0.25}$	&	$-3.94^{+0.13}_{-0.13}$	&	$-3.87^{+0.11}_{-0.10}$	&	$-3.83^{+0.10}_{-0.10}$	\\
$\ln\tau_{0}$	&	$6.4^{+2.2}_{-1.1}$	&	$5.3^{+0.4}_{-0.3}$	&	$5.2^{+0.2}_{-0.2}$	&	$5.3^{+0.2}_{-0.2}$	\\
$\gamma$	&	$1.14^{+0.17}_{-0.16}$	&	$1.44^{+0.12}_{-0.13}$	&	$1.51^{+0.09}_{-0.10}$	&	$1.50^{+0.08}_{-0.09}$	\\
EQUAD\_OVRO	&	$0.017^{+0.003}_{-0.004}$	&	$0.02^{+0.002}_{-0.003}$	&	$0.02^{+0.002}_{-0.002}$	&	$0.02^{+0.002}_{-0.002}$	\\
\enddata
\tablecomments{Posterior medians and $1\sigma$ credible intervals for circular SMBHB + DRW red noise model are reported. Note that the dashes (---) indicate that these parameters are not required for the corresponding dataset.}
\end{deluxetable*}

\begin{deluxetable*}{lcccc}
\tablecaption{Posterior estimates of our Eccentric SMBH binary + DRW red noise model for different datasets.\label{tab:posterior_eccDRW}}
\tablehead{
\colhead{Parameter} & \colhead{O} & \colhead{OH} & \colhead{OU} & \colhead{OHU}
}
\startdata
S0\_epoch1	&	---	&	$2.68^{+0.08}_{-0.08}$	&	$2.49^{+0.11}_{-0.11}$	&	$2.60^{+0.08}_{-0.08}$	\\
A\_epoch1	&	---	&	$0.88^{+0.10}_{-0.10}$	&	$0.65^{+0.14}_{-0.15}$	&	$0.81^{+0.11}_{-0.11}$	\\
EQUAD\_UMRAO	&	---	&	---	&	$0.10^{+0.01}_{-0.01}$	&	$0.10^{+0.01}_{-0.01}$	\\
EQUAD\_Haystack	&	---	&	$0.11^{+0.03}_{-0.03}$	&	---	&	$0.17^{+0.03}_{-0.03}$	\\
S0\_epoch3	&	$2.27^{+0.22}_{-0.11}$	&	$2.25^{+0.06}_{-0.05}$	&	$2.25^{+0.05}_{-0.05}$	&	$2.25^{+0.06}_{-0.05}$	\\
A\_epoch3	&	$0.35^{+0.07}_{-0.07}$	&	$0.36^{+0.07}_{-0.07}$	&	$0.34^{+0.06}_{-0.07}$	&	$0.33^{+0.07}_{-0.07}$	\\
$P_b$	&	$1684^{+104}_{-93}$	&	$1731^{+8}_{-7}$	&	$1734^{+12}_{-11}$	&	$1731^{+9}_{-9}$	\\
$t_0$	&	NaN	&	NaN	&	NaN	&	NaN	\\
$\ln\hat{\sigma}$	&	$-4.22^{+0.20}_{-0.25}$	&	$-3.94^{+0.13}_{-0.13}$	&	$-3.87^{+0.11}_{-0.10}$	&	$-3.83^{+0.10}_{-0.10}$	\\
$\ln\tau_{0}$	&	$6.5^{+2.2}_{-1.1}$	&	$5.2^{+0.3}_{-0.3}$	&	$5.2^{+0.2}_{-0.2}$	&	$5.3^{+0.2}_{-0.2}$	\\
$\gamma$	&	$1.14^{+0.17}_{-0.16}$	&	$1.44^{+0.12}_{-0.13}$	&	$1.51^{+0.09}_{-0.10}$	&	$1.51^{+0.09}_{-0.09}$	\\
EQUAD\_OVRO	&	$0.017^{+0.003}_{-0.004}$	&	$0.02^{+0.002}_{-0.003}$	&	$0.02^{+0.002}_{-0.002}$	&	$0.02^{+0.002}_{-0.002}$	\\
$\omega$	&	NaN	&	NaN	&	NaN	&	NaN	\\
$\log_{10}e (95\%)$	&	$-1.07$	&	$-1.14$	&	$-0.81$	&	$-0.86$	\\
\enddata
\tablecomments{Posterior medians and $1\sigma$ credible intervals for Eccentric SMBHB + DRW red noise model are reported. Note that the dashes (---) indicate that these parameters are not required for the corresponding dataset. $t_0$ and $\omega$ are highly covariant and are not constrained.}
\end{deluxetable*}

\begin{deluxetable*}{lcccc}
\tablecaption{Bayesian Evidence and Bayes Factors for Circular + DRW vs.\ Eccentric + DRW Models\label{tab:bayes_transposed_DRW}}
\tablehead{
\colhead{Quantity} & \colhead{O} & \colhead{OH} & \colhead{OU} & \colhead{OHU}
}
\startdata
$\ln Z$ (Circular + DRW)               & 1102.38 & 1092.69 & 1150.40 & 1142.31 \\
$\ln Z$ (Eccentric + DRW)                   & 1102.22 & 1092.49 & 1150.27 & 1142.16 \\
$\ln \mathcal{B}^{\rm ecc+DRW}_{\rm cir+DRW}$ & $-0.16$ & $-0.20$ & $-0.13$ & $-0.15$ \\
\enddata
\tablecomments{Comparison of natural log evidences ($\ln Z$) and natural log Bayes factors for the 
earlier circular and our eccentric ELL1 models across different data configurations.
Note that positive $\ln \mathcal{B}$ values suggest a preference for the ELL1 model.}
\end{deluxetable*}

\section{ Conclusions } \label{sec:conclusions}

Meticulous compilation of multi-wavelength data, especially radio light curves spanning many decades, intuitive theoretical modeling, and comprehensive data analysis efforts, detailed in \cite{pks_rb,PKS2025}, strongly supports a scenario where Blazar PKS 2131-021(z = 1.285) hosts an SMBH binary with an observed orbital period of $2.08$ yrs in its rest frame. The observed sinusoidal and  phase-stable radio variability could naturally be explained by the modulation of the strong Doppler boosting of the approaching relativistic jet, induced by the orbital motion of an SMBH binary \citep{pks_rb}. This theoretical description requires the constituents of SMBHs to move in circular orbits \citep{pks_rb}. Our effort incorporates analytically the effects of residual orbital eccentricity in the SMBH binary prescription for PKS 2131-021. We employ the Laplace-Lagrange parameters, widely employed to describe orbits of MSPs  with $e\ll1$ while using the \texttt{ELL1} timing model \citep{ELL1}. Our fully analytic approach describes the temporal variations in the radio flux density, induced by  SMBH binaries in tiny eccentric orbits, and is referred to as the ELL1 SMBH binary model. We provide arguments, based on time derivative of R{\o}mer Delay expression in \cite{ell1k}, to validate the correctness of our expression for $\delta \ln S$, given by Eq.~\ref{eq:ell1fluxchange}. Invoking radio light curves, employed in \cite{pks_rb}, we pursue detailed Bayesian parameter estimation investigations and 
conclude that the combined OVRO+Haystack+UMRAO (OHU) data show strong support for the presence of a tiny eccentricity in the SMBH binary description of PKS 2131‑021 ($e = 0.053 \pm 0.015$) when red noise is not considered in the analysis.
When the analysis is refined to account for the presence of red noise in our data through a Damped Random Walk (DRW) process — a stochastic model that captures the correlated variability intrinsic to quasar light curves — the circular orbital model emerges as statistically preferred, constraining the orbital eccentricity to an upper limit of e $<$ 0.15. Such eccentricity, persisting despite long-term orbital evolution, is a key dynamical signature that aligns with models of SMBH binary interactions in galactic nuclei \citep{Siwek2024,Rawlingseccsmbhb}. 
Nevertheless, the present effort reveals an important finding: the Circular+DRW model consistently recovers a coherent periodic signal across all datasets, with the orbital period remaining well-defined despite the broader uncertainties. These arguments further bolster the case for PKS 2131–021 as a credible source of nHz GWs \citep{pks_rb}.

\par 
It would be desirable to explore if our eccentric binary model is consistent with additional multi-wavelength data sets employed to substantiate the Kinematic Orbital model for PKS 2131–021 \citep{PKS2025,pks2131act,pks2131ren}. 
The current evidence for eccentricity is below the usual detection significance threshold of $\ln \mathcal{B} > 8$, and apsidal precession enters at higher order in the ELL1 SMBHB model. Therefore, a longer time baseline and more measurements are required to measure the eccentricity and the rate of apsidal precession reliably.
A possible effort could be related to the jet position angle (PA) of PKS 2131–021, and this refers to the angle made by the projected relativistic jet in the sky plane, usually measured from north through east. The temporal PA variations from the perspective of an SMBH binary central engine prescription for the unique blazar OJ~287 are explored systematically in \cite{Dey2021}, and it would be interesting to pursue a similar effort for PKS 2131–021. We believe that such investigations should establish PKS 2131–021 as a strong candidate for hosting an SMBH binary emitting in the nHz GW frequency window, alongside the well-studied unique blazar OJ~287. Its strong SMBH binary candidature is mainly due to the ability of OJ~287's SMBH binary central engine prescription to successfully predict future (observed) SMBH impact flares of 2005, 2007, 2015, and 2019 and how PA measurements should vary in the coming years \citep{Valtonen2016,Dey2018,Valtonen2021,Dey2021,Laine2020}.

\begin{acknowledgments}
We are grateful to  Vidit Singh and Adya Shukla for their invaluable assistance with high-performance computing. We thank Abhimanyu Susobhanan and José L. Gómez for helpful comments and discussions. The authors would like to extend their sincere gratitude to Sebastian Kiehlmann and  Margo Aller for providing the data on behalf of the OVRO 40m team. This research has made use of data from the OVRO 40-m monitoring program \citep{2011ApJS..194...29R}, supported by private funding from the California Institute of Technology and the Max Planck Institute for Radio Astronomy, and by NASA grants NNX08AW31G, NNX11A043G, and NNX14AQ89G and NSF grants AST-0808050, AST-1109911, AST-2407603 and  AST-2407604. AKP is supported by CSIR fellowship Grant number 09/0079(15784)/2022-EMR-I. AG acknowledges the support of the Department of Atomic Energy, Government of India, under project identification \#RTI 4002,
and the CAS President’s International Fellowship Initiative (2026PVA0020).

\end{acknowledgments}

\facilities{OVRO, UMRAO, Haystack Observatory}

\software{
\texttt{Nautilus} \citep{nautilus},
\texttt{SciPy} \citep{2020SciPy-NMeth,mckinney-proc-scipy-2010},
\texttt{corner} \citep{corner},
\texttt{Matplotlib} \citep{Hunter:2007},
\texttt{NumPy} \citep{harris2020array},
\texttt{pandas} \citep{reback2020pandas}
    }

\appendix

\section{Bayesian inference and model selection}
\label{sec:method}

Here, we describe the framework of Bayesian inference and model selection for the analysis of AGN light curve data, influenced by \cite{zhusuperbayes}. Assuming stationary Gaussian noise, the likelihood function is defined as
\begin{eqnarray}
\label{eq:likel}
\mathcal{L}(    {d} |    \Theta, \mathcal{H}) &= & \frac{1}{\sqrt{(2\pi)^{N} |\mathrm{C}|}} \exp\left[-\frac{1}{2} (   {d}-   {s})^T \mathrm{C}^{-1} (   {d}-   {s})\right] 
\end{eqnarray}
where $   {d}$ is the light curve data with length $N$ and $   \Theta$ include the model parameters for a given hypothesis $\mathcal{H}$. In this work, we use measurements of flux density with time. The data $   {d}$ is modeled as
\begin{equation}
\label{eq:data}
       {d}=   {s}+   {n}\, ,
\end{equation}
where $   {s}$ is the signal and $   {n}$ is the noise, which contains measurement uncertainties and additional intrinsic stochastic variability.

\par 
In Eq. \ref{eq:likel}, $\mathrm{C}_{ij}$ is the noise covariance matrix, assuming white noise it can be written as
\begin{equation}
\mathrm{C}_{ij, \mathrm{white}} = ( \sigma_{i}^2 + \sigma_{0}^2)\delta_{ij}\, ,
\label{eq:Cwhite}
\end{equation}
where $\sigma_{i}$ denotes the reported measurement uncertainty for the $i$th observation, and $\sigma_{0}$ is usually referred to as EQUAD \citep{zhusuperbayes}.
This is an additional variance term introduced to account for extra sources of noise and this contribution is added in quadrature to the measured uncertainties and serves to capture underestimated or unmodeled noise contributions, such as those arising from calibration errors, systematic trends, or intrinsic AGN variability on unresolved timescales. Incorporating $\sigma_{0}$ provides a more realistic estimate of the total observational uncertainty, ensuring that parameter inferences are robust against sources of variance not adequately reflected in the quoted error bars.
 
\par 
We use Bayesian model selection to quantify the statistical significance of the presence of eccentric signal in blazar light curve. We start with Bayes' theorem, which states that
\begin{equation}
P(   \Theta|   {d},\mathcal{H})= \frac{\mathcal{L}(   {d}|   \Theta,\mathcal{H})P(   \Theta|\mathcal{H})}{{\cal Z}(   {d}|\mathcal{H})}\, .
\end{equation}
Here $P(   \Theta|   {d},\mathcal{H})$ is the posterior probability distribution function of parameters $   \Theta$ given data $   {d}$ and hypothesis $\mathcal{H}$; $\mathcal{L}(   {d}|   \Theta,\mathcal{H})$ is the likelihood function given in equation (\ref{eq:likel}), which describes the probability of observing data given the hypothesis $\mathcal{H}$ and parameters $   \Theta$. Meanwhile, $P(   \Theta|\mathcal{H})$ is the prior distribution of parameters $   \Theta$ while ${\cal Z}(   {d}|\mathcal{H})$ is the Bayesian evidence for hypothesis $\mathcal{H}$
\begin{equation}
\label{eq:BayesZ}
{\cal Z}(   {d}|\mathcal{H}) = \int \text{d}   \Theta \mathcal{L}(   {d}|   \Theta,\mathcal{H})P(   \Theta|\mathcal{H})\, .
\end{equation}

Given the observational data, if we wish to compare two hypotheses: $\mathcal{H}_{1}$ and $\mathcal{H}_{2}$, we can use \textit{odds ratio} defined as the ratio of posterior probability:
\begin{equation}
\mathcal{O}=\frac{P(\mathcal{H}_{1}|   {d})}{P(\mathcal{H}_{2}|   {d})} = \frac{\mathcal{Z}(   {d}|\mathcal{H}_{1})P(\mathcal{H}_{1})}{\mathcal{Z}(   {d}|\mathcal{H}_{2})P(\mathcal{H}_{1})}\, ,
\end{equation}
where $P(\mathcal{H}_{1})$ and $P(\mathcal{H}_{2})$ are the prior probability for hypotheses $\mathcal{H}_{1}$ and $\mathcal{H}_{2}$, respectively. Assuming equal prior probability for both hypotheses, Bayesian model selection is usually performed by computing the Bayes factor:
\begin{equation}
\mathcal{B}^{\rm 1}_{\rm 2} = \frac{\mathcal{Z}(   {d}|\mathcal{H}_{1})}{\mathcal{Z}(   {d}|\mathcal{H}_{2})}\, .
\end{equation}
In this effort, we are interested to explore the support for our eccentric model quantified by $\mathcal{B}^{\rm 1}_{\rm 2}$ and we invoke \cite{rafterybayesfac} for the interpretation of Bayes factors as summarized in Table \ref{tab:bayes_interpretation}. 

\begin{table}[H]
\centering
\caption{ A guide for interpreting Natural Log Bayes Factors}
\label{tab:bayes_interpretation}
\begin{tabular}{lc}
\hline
\hline
Range of $\ln \mathcal{B}^{\rm 1}_{\rm 2}$ & Interpretation \\
\hline
$0 < \ln \mathcal{B} < 1$   & Worth no more than a bare mention \\
$1 < \ln \mathcal{B} < 3$   & Positive support \\
$3 < \ln \mathcal{B} < 5$   & Strong support \\
$\ln \mathcal{B} > 5$       & Very strong support \\
$\ln \mathcal{B} > 8$       & Detection threshold used in GW studies \\
\hline
\end{tabular}
\tablecomments{ These interpretations are based on \citet{rafterybayesfac} 
and we note that a threshold of $\ln \mathcal{B} > 8$ is often used in GW astronomy as evident from  \cite{thranegwbayes}.}
\end{table}

\section{Fractional Uncertainty in SINE and COSINE of True Anomaly}
\label{app:uncertainty}
To determine the point at which the ELL1 SMBHB model approximation breaks down, we computed the mismatch metric, $m$, defined as:
\begin{equation}
    m = 1 - \frac{aN^{-1}b}{\sqrt{(aN^{-1}a)(bN^{-1}b)}}
\end{equation}
where
\begin{equation}
    a = y-y_{approx}
\end{equation}
\begin{equation}
    b = y-y_{true}
\end{equation}
\begin{equation}
    N = diag(\sigma^2 + EQUAD^2)
\end{equation}
Here, $y$ is the data, $y_{approx}$ is the model with ELL1 approximation and $y_{true}$ is the model obtained by solving the kepler's equation. $\sigma$ and $EQUAD$ represents the uncertainties corresponding to $y$. We computed the mismatch metric with varying eccentricity and keeping other parameters constant. By adopting a threshold $ m = 10^{-3}$ as the limit, we found ELL1 SMBHB model is valid for $e <0.12$.

\bibliography{sample701}{}
\bibliographystyle{aasjournal}

\end{document}